\pdfoutput=1

\documentclass[aps,10pt,twocolumn,showpacs,superscriptaddress,footnoteinbib]{revtex4}
\usepackage{amsmath}
\usepackage{latexsym}
\usepackage{amssymb}
\usepackage{graphics,epstopdf}
\usepackage[colorlinks=true, citecolor=blue, urlcolor=blue ]{hyperref}

\begin{document}
\title{Maximum tri-partite Hardy's nonlocality respecting all bi-partite principles}
\author{Subhadipa Das}
\email{sbhdpa.das@bose.res.in}
\affiliation{S.N. Bose National Center for Basic Sciences, Block JD, Sector III, Salt Lake, Kolkata-700098, India}

\author{Manik Banik}
\email{manik11ju@gmail.com}
\affiliation{Physics and Applied Mathematics Unit, Indian Statistical Institute, 203 B.T. Road, Kolkata-700108, India}

\author{MD Rajjak Gazi}
\email{rajjakgazimath@gmail.com}
\affiliation{Physics and Applied Mathematics Unit, Indian Statistical Institute, 203 B.T. Road, Kolkata-700108, India}

\author{Ashutosh Rai}
\email{arai@bose.res.in}
\affiliation{S.N. Bose National Center for Basic Sciences, Block JD, Sector III, Salt Lake, Kolkata-700098, India}

\author{Samir Kunkri}
\email{skunkri@yahoo.com}
\affiliation{Mahadevananda Mahavidyalaya, Monirampore, Barrakpore, North 24 Parganas-700120, India}

\author{Ramij Rahaman}
\email{ramij.rahaman@ug.edu.pl}
\affiliation{The Institute of Mathematical Sciences (IMSc.), CIT Campus, Taramani, Chennai-600113, India}
\affiliation {Institute of Theoretical Physics and Astrophysics, University of Gdansk, 80-952 Gdansk, Poland}

\begin{abstract}
The set of multiparty correlations that respect all bi-partite principles has been conjectured to be same as the set of time-ordered-bi-local correlations. Based on this conjuncture we find the maximum value of success probability of tri-partite Hardy's correlation respecting all bi-partite physical principles. Unlike in quantum mechanics, the no-signaling principle does not reveal any gap in Hardy's maximum success probability for bi-partite and tri-partite system. Information causality principle is shown to be successful in qualitatively revealing this quantum feature and this result is independent of the conjecture mentioned above.
\end{abstract}

\maketitle

\section{Introduction}
The outcome of local measurements on spatially separated parts of a composite quantum system can be non-classically (nonlocally) correlated. Violation of the Bell-CHSH inequality \cite{bell,chsh} is a witness of this nonlocal feature in such correlations. The value of Bell-CHSH expression, exceeding the classical bound $2$, then qualifies as a measure of nonlocality. This nonlocality within the quantum mechanics is limited by the Cirel'son bound $2\sqrt{2}$ \cite{cirel}. On the other hand, on considering a larger set of generalized correlations (possibly non-quantum) which are compatible with \emph{no signaling} (NS) principle, nolocality underlying these correlations can achieve any value up to the algebraic maximum $4$ for the Bell-CHSH expression (e.g. PR-box correlation achieves the value $4$ \cite{popescu}). So the natural questions arises; what are the physical principles, other than NS, that can distinguish quantum correlations
from post-quantum no-signaling correlations? This fundamental question has been addressed in several recent works proposing novel physical principles, like, no nontrivial communication complexity \cite{vandam1,vandam2}, macroscopic locality \cite{navas} and information causality \cite{pawlowski}, for explaining the boundary defining quantum correlations. In particular, the application of \emph{the principle of noviolation of information causality} (IC) has produced very interesting results, like explaining the Cirel'son's bound and showing that in a bipartite scenario any correlation going beyond the Cire'lson bound is unphysical. IC principle is a generalization of no-signaling condition---while relativistic causality (the no-signaling principle) states that a party cannot extract more information then the communicated (say, $m$) number of cbits; information causality further restricts \emph{free choice to decode deterministically} a single $m$-cbit information, from different possible $m$-cbit informations potentially encoded within the communicated amount of $m$-cbit message. Applications of the IC principle in the study of both bi-partite and tri-partite correlations has produced some interesting results \cite{allcock,ali,gazi,yang}.
On the other hand, IC or any other bi-partite principle has been shown to be insufficient for witnessing all multipartite post-quantum correlations \cite{gallego,subhadipa}. Thus, multipartite generalization of IC, or some other genuinely multipartite physical principle(s) are necessary to characterize all quantum correlations. Studying some simple class of multipartite correlations, like Hardy's set \cite{hardy,hardy1,samir1,rabelo,ghosh}, can give useful insight about the strength and weakness of such principles \cite{ali,gazi,subhadipa}.

Like Bell-CHSH nonlocality test, Lucian Hardy first proposed \cite{hardy} an elegant argument for witnessing nonlocal correlations without any use of inequality. For two qubit states subjected to local projective measurements, the maximum success probability of Hardy's nonlocality argument has been shown to be $(5\sqrt{5}-11)/2$ ($\approx0.09$)\cite{hardy1,samir1}. Recently, it has also been shown that this value is `device independent', i.e, the maximum success probability of Hardy's argument for any bipartite quantum state is $(5\sqrt{5}-11)/2$ \cite{rabelo}. Again, on extending Hardy's argument to three qubit systems subject to local projective measurements, the maximum success probability of the argument is shown to be  $\frac{1}{8}~(=0.125)$  \cite{ghosh}---this value also holds for any tripartite state subjected to arbitrary measurements, i.e., this value is also device independent \cite{subhadipa}. Thus, in quantum mechanics, the maximum success probability of Hardy's nonlocality argument for tri-partite system is greater than that for bi-partite system.

Hardy's nonlocality has been studied in the broader framework of general probabilistic theories by invoking physical principles like NS condition and IC condition. Under the NS condition, optimal success probability for Hardy's nonlocality is $\frac{1}{2}$, both for bi-partite system and tri-partite system \cite{ghosh}. Thus, in contrast to the quantum mechanical feature, under the NS condition there is no gap between Hardy's maximum success probability for bi-partite and tri-partite system. On the other hand, under the IC principle, it has been shown that the maximum success probability of Hardy's argument for bi-partite system is bounded above by $0.207$ \cite{ali}. The study for the bound on Hardy's success probability for tri-partite system under IC condition has not yet been studied. The problem is very intriguing as information causality is a bi-partite principle and it is highly nontrivial to exhaust the IC condition under all bi-partitions with all possible wirings.

In this work we show that the maximum value of Hardy's success for tri-partite correlation satisfying every bi-partite principle is $\frac{1}{4}$. Then, we argue that in particular IC principle successfully reveals a quantum feature viz. a gap between Hardy's maximum success probability for bi-partite and tri-partite systems. Moreover, the gap between two bounds is decisive, as for tri-partite system we achieve a lower bound through a probability distribution which is \emph{time-ordered-bi-local} (TOBL) \cite{pironio,gallego1,barrett} and hence it not only satisfies IC but satisfies  any bi-partite information principles discovered or not discovered. On the other hand, for the bipartite case, the upper bound on maximum success probability was derived by applying a necessary condition for respecting the IC principle.

The paper is organized as follows. In section (\ref{nosignaling}) we overview the properties of the sets of a tripartite two input two output probability distributions. In section (\ref{hardy}), we describe Hardy's non-locality conditions, and review some results in this context. In section (\ref{tobl}), we briefly describe TOBL correlations. In section (\ref{hardyic}), we derive that maximum success of tri-partite Hardy's nonlocality respecting all bi-partite information principle, and then in section (\ref{ic}), we argue that, in the context of maximum Hardy's success probability, IC reproduces a quantum like feature while NS fails to do so. Finally we give our conclusions in section (\ref{conclusion}).

\section{Tripartite No-signaling Correlations}\label{nosignaling}
The set of tripartite no-signaling correlations with binary input and binary output for each party is a convex set in a 26-dimensional space \cite{pironio}. Let $P(abc|xyz)$ denotes the tri-joint probabilities, where $x,y,z\in\{0,1\}$ denote inputs and $a,b,c\in\{0,1\}$ denote outputs of three parties respectively. Normalization and positivity of joint probabilities are expressed by following conditions:
\begin{eqnarray}
\sum_{abc}P(abc|xyz)=1~~\forall ~x,y,z,\\
P(abc|xyz)\geq0~~\forall~ a,b,c,x,y,z.
\end{eqnarray}
The set of no-signaling distributions is obtained by further imposing the condition that the choice of measurement by any one party cannot affect the outcome distributions of the remaining two parties and vice-versa, i.e.,
\begin{equation}
 \sum_{c}p(abc|xyz)=\sum_{c}p(abc'|xyz'),~~\forall a,b,x,y,z,z'
\end{equation}
and other similar conditions obtained by permuting the parties.
The set of these no-signaling correlations forms a polytope with $53,856$ extremal points belonging to $46$ different classes among which $45$ contains non-local points \cite{pironio}. The remaining class contains all deterministic local points and correspond to extreme points of the local polytope; i.e., the set of correlations with a local model
\begin{equation}
 p(abc|xyz)=\sum_{\lambda}p_{\lambda}p(a|x,\lambda)p(b|y,\lambda)p(c|z,\lambda)
\end{equation}
where $p_{\lambda}$ is the distribution of some random variable $\lambda$ shared by the parties.
Since, in general quantum correlations respects the no-signaling condition, set of all tripartite quantum correlations with two possible outcomes at each local site are obviously contained within the tripartite no-signaling polytope.
Then, a correlation is quantum if it can be expressed as
\begin{equation}
 p(abc|xyz)= tr(\rho M_x^a\otimes M_y^b\otimes M_z^c)
\end{equation}
where $\rho$ is some quantum state and indexed $M$ are the measurement operators (in general positive operator valued measure POVM) associated with the outcomes $a,b,c$ of measurements $x,y,z$ respectively. Post-quantum no-signaling correlations are those that cannot be written in the above forms (i.e. eqn-(4) or (5)).

\section{Tripartite Hardy's correlations}\label{hardy}
A tripartite two-input-two-output Hardy's correlation is defined by some restrictions on a certain choice of $5$ out of $64$ joint probabilities in the correlation matrix. Any tripartite Hardy's correlation can be expressed by following five conditions:
\begin{subequations}
\begin{eqnarray}
 P(A=i, ~B=j, ~C=k)=q_3 > 0\\
 P(A'=l, ~B=m, ~C~=k)=0\\
 P(A=i, ~B'=m, ~C=k)=0\\
 P(A=i, ~B=j, ~C'=n)=0 \\
 P(A'=l\oplus1,~B'=m\oplus1,~C'=n\oplus1)=0
\end{eqnarray}
\end{subequations}
where, $\{A,~A'\}$, $\{B,~B'\}$ and $\{C,~C'\}$ are the respective outcomes for the choice of local observables for three parties, say, Alice, Bob and Charlie, and $i,j,k,l,m,n \in \{0,1\}$.

One can prove that Hardy's correlations are non-local. To show this, let us suppose that these correlations are local, i.e., these correlations can be simulated from pre-shared (hidden) local variables. Now, consider the subset of those hidden variables $\lambda$ shared between the three parties such that the first condition holds. Then, the second, the third, and the fourth condition, together with the first condition implies that $P(A'=l\oplus1,~B'=m\oplus1,~C'=n\oplus1)>0$ in contradiction with the last (fifth) condition. Hence Hardy's correlation must be non-local. Also note that, $q_3$, the value of the joint probability appearing in the first condition is the success probability of the Hardy's argument. In quantum mechanics, for tri-partite system the maximum value of $q_3$ is $0.125$ \cite{subhadipa}. On the other hand in a general no-signaling theory $q_3$ attains a maximum value $0.5$ \cite{ghosh}.

Known results for the success probability, say $q_2$, of a bipartite two-input-two-output Hardy's argument are: (i) Under no-signaling constraint $max(q_2)=0.5$, which is same as for the tripartite correlations \cite{ghosh}, (ii) Under the information causality principle $max(q_2)$ is bounded from above by $0.207$ \cite{ali}, and (iii) In quantum mechanics, $max(q_2)=0.09$ \cite{rabelo}.


\section{Time-ordered Bi-Local Correalions}\label{tobl}
A tripartite no-signaling probability distribution  $P(abc|xyz)$ belongs to TOBL \cite{pironio,gallego1,barrett} if it can be written as
\begin{eqnarray}
P(abc|xyz)=\sum_{\lambda}p_{\lambda}P(a|x,\lambda)P_{B \rightarrow C}(bc|yz,\lambda)\\
          =\sum_{\lambda}p_{\lambda}P(a|x,\lambda)P_{B \leftarrow C}(bc|yz,\lambda)
\end{eqnarray}
and analogously for $B|AC$ and $C|AB$, where $p_{\lambda}$ is the distribution of some random variable $\lambda$, shared by the parties. The distributions $P_{B \rightarrow C}$ and $P_{B \leftarrow C}$ respect the conditions
\begin{eqnarray}
P_{B \rightarrow C}(b|y,\lambda)=\sum_{c}P_{B \rightarrow C}(bc|yz,\lambda)\\
P_{B \leftarrow C}(c|z,\lambda)=\sum_{b}P_{B \leftarrow C}(bc|yz,\lambda)
\end{eqnarray}
From these equations it is clear that the distributions $P_{B \rightarrow C}$ allow signaling from Alice to Bob and $P_{B \leftarrow C}$ allow signaling from Bob to Alice.
If a tripartite no-signaling probability distribution $P(abc|xyz)$ belongs to the set of TOBL distributions, all possible bipartite distributions derived by applying any local wirings on $P(abc|xyz)$ are local, i.e., the probability distribution $P(abc|xyz)$ in general respects any bi-partite physical principles, therefore, any TOBL correlation must also respect the information causality principle.

\section{Maximum tri-partite Hardy's nonlocality respecting all bi-partite principles}\label{hardyic}
Without loss of generality we consider following tri-partite Hardy's correlation:
\begin{subequations}
\begin{eqnarray}
P(110|001)> 0 \\
P(110|000)=0\\
P(100|011)=0\\
P(010|101)=0\\
P(111|110)=0
\end{eqnarray}
\end{subequations}
It is known that any tripartite no-signaling probability distribution $P(abc|xyz)$ belonging to TOBL set respects all bi-partite information principles. On maximizing success probability $P(110|001)$ (by a program written in MATHEMATICA) over the set of TOBL correlations, the achieved maximum value is $\frac{1}{4}$. A probability distribution which attains this maximum value $\frac{1}{4}$ for Hardy nonlocality is given in TABLE \ref{TABLE I}.
\begin{table}[t]
\caption{Tripartite no-signaling probability distribution
$ P(abc|xyz)$ with maximum Hardy's success $1/4$.}
\begin{tabular}{lllllllll}
\hline\hline
$xyz\backslash abc| $ & $000$ & $001$ & $010$ & $011$ & $100$ & $101$ &$110$ & $111$ \\
\hline
~~$000$ &~$\frac{1}{4}$ &~ $0$ & ~$0$ & ~$\frac{1}{4}$ &~ $\frac{1}{4}$ &~ $0$ &~ $0$ & ~$\frac{1}{4}$ \\
~~$001$ &~$\frac{1}{4}$ &~ $0$ & ~$0$ & ~$\frac{1}{4}$ &~ $0$ & ~$\frac{1}{4}$ &~ $\frac{1}{4}$ & ~$0$ \\
~~$010$ &~$\frac{1}{4}$ &~ $\frac{1}{4}$ & ~$0$ & ~$0$ &~ $0$ &~ $0$ &~ $\frac{1}{4}$ & ~$\frac{1}{4}$ \\
~~$011$ &~$\frac{1}{4}$ &~ $\frac{1}{4}$ & ~$0$ & ~$0$ &~ $0$ &~ $0$ &~ $\frac{1}{4}$ & ~$\frac{1}{4}$ \\
~~$100$ &~$\frac{1}{4}$ &~ $0$ & ~$0$ & ~$\frac{1}{4}$ &~ $\frac{1}{4}$ &~ $0$ &~ $0$ & ~$\frac{1}{4}$ \\
~~$101$ &~$0$ &~ $\frac{1}{4}$ & ~$0$ & ~$\frac{1}{4}$ &~ $\frac{1}{4}$ &~ $0$ &~ $\frac{1}{4}$ & ~$0$ \\
~~$110$ &~$\frac{1}{4}$ &~ $0$ & ~$0$ & ~$\frac{1}{4}$ &~ $0$ &~ $\frac{1}{4}$ &~ $\frac{1}{4}$ & ~$0$ \\
~~$111$ &~$0$ &~ $\frac{1}{4}$ & ~$0$ & ~$\frac{1}{4}$ &~ $\frac{1}{4}$ &~ $0$ &~ $\frac{1}{4}$ & ~$0$ \\
\hline\hline
\end{tabular}
\label{TABLE I}
\end{table}

We now construct a TOBL model (TABLES (\ref{TABLE II}-\ref{TABLE VII})) for the distribution $P(abc|xyz)$ given in TABLE \ref{TABLE I}.
Probability distribution appearing in the TOBL decomposition for the bi-partition $A|BC$, are such that for a given $\lambda$ Alice's outcome $a$ depends only her measurement settings $x$. Also, for given $\lambda$, $P_{B \rightarrow C}(b|y,\lambda)$ is independent of $z$ but for $B  \rightarrow C$, $c$ depends on both $ y$ and $z$. Similarly, for given $\lambda$, $P_{B\leftarrow C}(c|z,\lambda)$ is independent of $y$ but for $B\leftarrow C$, $b$ depends on both $ y$ and $z$. Let $a_x$, $b_y$ and $c_{z}$ denote the outcomes for Alice, Bob and Charlie for the respective inputs $x$, $y$ and $z$. For all TABLES (\ref{TABLE II}-\ref{TABLE VII}) in TOBL decompositions, the outputs are deterministic and the weights $p_\lambda$ are same. And the outcome assignments of $A$ for $A|B \rightarrow C$ and $A|B \leftarrow C$ are same and similar is true for other two bipartition. We summarize the above result in the following proposition.

{\bf Proposition:} \emph{The maximum success probability for Hardy's nonlocality argument for tri-partite time-ordered-bi-local (TOBL) correlations is $\frac{1}{4}$.}

Now, in view of the conjuncture made in \cite{gallego1}, TOBL set constitute the largest set of correlations that remains consistent under \emph{wirings and classical communication prior to the inputs} protocols when some parties collaborate (i.e., the set of correlations respecting all bi-partite principles). Based on this conjuncture, the value $\frac{1}{4}$ in the above proposition is the maximum value of Hardy's nonlocality for tri-partite correlations satisfying all bi-principles.

\begin{table}[t]
\caption{TOBL model $A|B \rightarrow C$.}
\begin{tabular}{lllllllllll}
\hline
$ $ & $\lambda$ & $p_\lambda $ & $a_0$ & $a_1$ & $b_0$ & $b_1$ & $c_{00}$ & $c_{01}$ & $c_{10}$ & $c_{11}$ \\
\noalign{\smallskip}\hline\noalign{\smallskip}
$ $ & $1$ & $\frac{1}{4}$ & $0$ & $0$ & $1$ & $0$ & $1$ & $1$ & $0$ & $1$ \\
$ $ & $2$ & $\frac{1}{4}$ & $1$ & $0$ & $0$ & $1$ & $0$ & $1$ & $1$ & $1$ \\
$ $ & $3$ & $\frac{1}{4}$ & $0$ & $1$ & $0$ & $0$ & $0$ & $0$ & $1$ & $0$ \\
$ $ & $4$ & $\frac{1}{4}$ & $1$ & $1$ & $1$ & $1$ & $1$ & $0$ & $0$ & $0$ \\
\hline
\end{tabular} \label{TABLE II}
\caption{ TOBL model $A|B\leftarrow C$.}
\begin{tabular}{lllllllllll}
\hline
$ $ & $\lambda $ & $p_\lambda$ & $a_0$ & $a_1$ & $b_{00}$ & $b_{01}$ & $b_{10}$ & $b_{11}$ & $c_{0}$ & $c_{1}$ \\
\noalign{\smallskip}\hline\noalign{\smallskip}
$ $ & $1$ & $\frac{1}{4}$ & $0$ & $0$ & $0$ & $1$ & $0$ & $0$ & $0$ & $1$ \\
$ $ & $2$ & $\frac{1}{4}$ & $1$ & $0$ & $1$ & $0$ & $1$ & $1$ & $1$ & $1$ \\
$ $ & $3$ & $\frac{1}{4}$ & $0$ & $1$ & $1$ & $0$ & $0$ & $0$ & $1$ & $0$ \\
$ $ & $4$ & $\frac{1}{4}$ & $1$ & $1$ & $0$ & $1$ & $1$ & $1$ & $0$ & $0$ \\
\hline
\end{tabular}
\label{TABLE III}
\caption{TOBL model $B|A \rightarrow C$.}
\begin{tabular}{lllllllllll}
\hline
$ $ & $\lambda $ & $p_\lambda$ & $b_0$ & $b_1$ & $a_0$ & $a_1$ & $c_{00}$ & $c_{01}$ & $c_{10}$ & $c_{11}$ \\
\noalign{\smallskip}\hline\noalign{\smallskip}
$ $ & $1$ & $\frac{1}{4}$ & $0$ & $0$ & $0$ & $0$ & $0$ & $0$ & $0$ & $1$ \\
$ $ & $2$ & $\frac{1}{4}$ & $1$ & $0$ & $0$ & $1$ & $1$ & $1$ & $1$ & $0$ \\
$ $ & $3$ & $\frac{1}{4}$ & $0$ & $1$ & $1$ & $1$ & $0$ & $1$ & $0$ & $0$ \\
$ $ & $4$ & $\frac{1}{4}$ & $1$ & $1$ & $1$ & $0$ & $1$ & $0$ & $1$ & $1$ \\
\hline
\end{tabular}
\label{TABLE IV}
\caption{TOBL model $B|A \leftarrow C$.}
\begin{tabular}{lllllllllll}
\hline
$ $ & $\lambda $ & $p_\lambda$ & $b_0$ & $b_1$ & $a_{00}$ & $a_{01}$ & $a_{10}$ & $a_{11}$ & $c_0$ & $c_1$ \\
\noalign{\tiny}\hline\noalign{\tiny}
$ $ & $1$ & $\frac{1}{4}$ & $0$ & $0$ & $0$ & $0$ & $0$ & $1$ & $0$ & $0$ \\
$ $ & $2$ & $\frac{1}{4}$ & $1$ & $0$ & $0$ & $0$ & $1$ & $0$ & $1$ & $1$ \\
$ $ & $3$ & $\frac{1}{4}$ & $0$ & $1$ & $1$ & $1$ & $1$ & $0$ & $0$ & $1$ \\
$ $ & $4$ & $\frac{1}{4}$ & $1$ & $1$ & $1$ & $1$ & $0$ & $1$ & $1$ & $0$ \\
\hline
\end{tabular}
\label{TABLE V}


\caption{TOBL model $A \rightarrow B|C$.}
\begin{tabular}{lllllllllll}
\hline
$ $ & $\lambda $ & $p_\lambda$ & $a_0$ & $a_1$ & $b_{00}$ & $b_{01}$ & $b_{10}$ & $b_{11}$ & $c_0$ & $c_1$ \\
\noalign{\tiny}\hline\noalign{\tiny}
$ $ & $1$ & $\frac{1}{4}$ & $0$ & $1$ & $0$ & $0$ & $0$ & $1$ & $0$ & $0$ \\
$ $ & $2$ & $\frac{1}{4}$ & $1$ & $1$ & $1$ & $1$ & $1$ & $0$ & $1$ & $0$ \\
$ $ & $3$ & $\frac{1}{4}$ & $1$ & $0$ & $0$ & $1$ & $0$ & $0$ & $0$ & $1$ \\
$ $ & $4$ & $\frac{1}{4}$ & $0$ & $0$ & $1$ & $0$ & $1$ & $1$ & $1$ & $1$ \\
\hline
\end{tabular}
\label{TABLE VI}
\caption{TOBL model $A \leftarrow B|C$.}
\begin{tabular}{lllllllllll}
\hline
$ $ & $\lambda $ & $p_\lambda$ & $a_{00}$ & $a_{01}$ & $a_{10}$ & $a_{11}$ & $b_0$ & $ b_1$ & $c_0$ & $c_1$ \\
\noalign{\tiny}\hline\noalign{\tiny}
$ $ & $1$ & $\frac{1}{4}$ & $0$ & $1$ & $1$ & $1$ & $0$ & $1$ & $0$ & $0$ \\
$ $ & $2$ & $\frac{1}{4}$ & $1$ & $0$ & $1$ & $1$ & $1$ & $0$ & $1$ & $0$ \\
$ $ & $3$ & $\frac{1}{4}$ & $1$ & $0$ & $0$ & $0$ & $0$ & $0$ & $0$ & $1$ \\
$ $ & $4$ & $\frac{1}{4}$ & $0$ & $1$ & $0$ & $0$ & $1$ & $1$ & $1$ & $1$ \\
\hline
\end{tabular}
\label{TABLE VII}
\end{table}


\section{Quantum Like Feature UNDER IC}\label{ic}
When two distance observers are involved, IC principle has been proved to be more efficient to single out the physically allowed correlations in comparison to NS principle. But when more than two distance observers are involved both IC and NS principles have been proved to be insufficient to witness physical correlations \cite{gallego,subhadipa,yang}. However, here we argue that IC, in spite of being a bi-partite principle like NS, exhibits a qualitative quantum like feature when set of multipartite correlations are considered. Interestingly, this feature cannot be reproduced just by considering the NS principle.

According to our proposition stated in the section (\ref{hardyic}), as the maximum success probability of Hardy's argument for tri-partite system satisfying any bi-partite information principle can not be less than $\frac{1}{4}$, it is even more true when IC principle is the only constraint. On the other hand, for bi-partite system, applying a necessary condition for respecting the IC principle, the maximum success probability of Hardy's nonlocality argument have been shown to be bounded above by $0.207$ \cite{ali}. Therefore, there is a gap between maximum success probabilities of Hardy's nonlocality argument for bi-partite system and tri-partite system under the constraint of IC principle and this gap is decisive in the sense that it may increase but cannot decrease. If we consider the scenario within quantum correlations then from the results of \cite{rabelo} and \cite{subhadipa} we observed a genuine gap between maximum success probabilities for Hardy's nonlocality argument for bi-partite quantum system and tri-partite quantum system (in \cite{rabelo} (\cite{subhadipa}) it has been shown that maximum success probabilities for Hardy's nonlocality argument for bi(tri)-partite system is $0.09$($\frac{1}{8}$) ). Interestingly, the set of NS correlations both for bi-partite and tri-partite scenario exhibit Hardy's nonlocality argument with same maximum success probability $0.05$--thus in this case the gap, shown by quantum correlations as well as IC respecting correlations, is not observed.


\section{Conclusion}\label{conclusion}
Time-ordered-bi-local correlation is an important characterization to study nonlocality in a multiparty scenario. We studied the tripartite Hardy's nonlocality by spanning the set of all TOBL correlations. We obtain the maximum value of Hardy's success probability within this set to be $\frac{1}{4}$. Assuming the conjuncture made in \cite{gallego1} to be true, this value is the maximum success probability of tripartite Hardy's correlation respecting all bi-partite principles. In case, if the conjuncture does not hold the derived value $\frac{1}{4}$ is a lower bound on the maximum success probability for tripartite Hardy's correlation satisfying all bi-partite principles.
We argue that though IC principle can not reproduce various quantum features quantitatively like maximum success probability for Hardy's nonlocality in quantum mechanics, still it could reproduce a qualitative interesting quantum feature like revealing a gap between bi-partite and tri-partite cases for the maximum success probability of Hardy's nonlocality argument; this result holds independent of the above mentioned conjuncture. Thus though IC, being a bi-partite principle, is insufficient for the reproduction of physically allowed correlations for multipartite case \cite{gallego}, it can still be a useful in reproducing many quantum like features even when three or more spatially separated observers are involved.

\begin{acknowledgments}
It is a pleasure to thank Guruprasad Kar for many stimulating discussion. SK acknowledges Sibasish Ghosh for fruitful discussions during his recent visit to IMSC. SD and AR acknowledges support from the DST project SR/S2/PU-16/2007. RR acknowledges partial support by TEAM program of Foundation for Polish Science and ERC grant QOLAPS.
\end{acknowledgments}


\end{document}